\documentclass[1p]{elsarticle}

\usepackage{subcaption}
\usepackage{tikz}
\usepackage{pgfplots}
\usetikzlibrary{trees,positioning,arrows,decorations.pathreplacing, shapes,calc, external}
\usepackage{color}
\usepackage{amssymb}
\usepackage{amsmath, cases}
\usepackage{booktabs}
\usepackage{lineno,hyperref}

\makeatletter
\def\ps@pprintTitle{%
	\let\@oddhead\@empty
	\let\@evenhead\@empty
	\def\@oddfoot{}%
	\let\@evenfoot\@oddfoot}
\makeatother

\newdefinition{rmk}{Remark}
\newproof{pf}{Proof}

\biboptions{authoryear}

\begin{document}
	
	\begin{frontmatter}
		
		\title{Identifying Influential Nodes in Bipartite Networks Using the Clustering Coefficient} 
		
		\author[mymainaddress]{J. Liebig} 
		\author[mymainaddress]{A. Rao}
		\address[mymainaddress]{School of Mathematical and Geospatial Sciences, RMIT University, Melbourne 3001, Australia}
		
		\begin{abstract}
			The identification of influential nodes in complex network can be very challenging. If the network has a community structure, centrality measures may fail to identify the complete set of influential nodes, as the hubs and other central nodes of the network may lie inside only one community. Here we define a bipartite clustering coefficient that, by taking differently structured clusters into account, can find important nodes across communities. 
		\end{abstract}
		
		
	\end{frontmatter}

\section{Introduction}\label{sec:intro}
Locating important nodes in a network is often crucial as this could aid in terminating the spread of diseases or alternatively assist the spread of knowledge and information \citep{che12}. A number of centrality measures are currently used to identify important nodes, but as \citet{kit10} point out, these measures may not reveal the truly important nodes. This is especially the case if the network has a community structure, where centrality measures may only reveal important nodes from one of the communities \citep{zha13}. This paper uses a very different approach, defining new clustering coefficients and using these to find influential nodes across communities.  

Here we focus on bipartite networks. Many real world systems are best modelled as such. Examples are collaboration networks \citep{new01}, roosting spatial networks \citep{for09} and board interlocks \citep{pie13}. Networks are bipartite if their nodes can be partitioned into two disjoint sets (primary and secondary) such that the edges only connect nodes from different sets \citep{asr98}. In a collaboration network, for instance, the authors are only connected to papers and form the primary set. The papers form the secondary set. 

In order to analyse bipartite networks it is very common to study a one-mode mapping of the original bipartite network. This approach is called one-mode projection \citep{was94} or `conversion' \citep{bor11}. The bipartite network is projected onto a one-mode network by dropping one of the two node sets and connecting two nodes in the one-mode network if they share a neighbour in the bipartite network. This popular approach is necessitated by the fact that many network measures cannot be directly applied to bipartite networks \citep{lat08}. However, the projection onto a one-mode network leads to loss of information \citep{con12, vog10, zho07} and more recently many measures have been specifically redefined to suit the analysis of bipartite networks \citep{bor12}. 

In this paper we work directly on the bipartite network, defining a clustering coefficient for the analysis. We show that clusters in bipartite networks may have different structures and that ignoring these leads to inaccurate results. We then use our measure to identify nodes that drive the clustering behaviour of the network and show that these are indeed influential nodes.  

The rest of the paper is organised as follows: Section \ref{sec:preMethods} discusses previous definitions of the bipartite clustering coefficient and identifies their limitations. Section \ref{sec:formation} shows that differently structured bipartite clusters have distinct origins that depend on the way in which the network develops over time. In Section \ref{sec:improvedClustering} we give a bipartite clustering coefficient that can be used to identify important nodes. The proposed  method that finds important nodes is outlined in Section \ref{sec:importantNodes}. Sections \ref{sec:women} and \ref{sec:Noordin} discuss results that are obtained from applying our clustering coefficient to real world data sets. The results give insight into how clusters are structured and reveal the nodes that drive the formation of clusters and their particular structure.

\section{The Bipartite Clustering Coefficient}\label{sec:preMethods}
The clustering coefficient is one particular measure that, as originally defined, cannot be applied to bipartite networks. In a one-mode network, the clustering coefficient measures the concentration of triangles. It is an important measure in the analysis of networks, since it gives insight into how well the neighbourhood of a node is connected. As bipartite networks are triangle free, this measure cannot be directly applied to these networks. Several definitions of the bipartite clustering coefficient exist \citep{lin05, ops13, rob04, zha08}. They are, however, inconsistent and hence require further investigation. 

Most existing bipartite clustering coefficients measure the concentration of 4-cycles instead of triangles (\citep{lin05, rob04, zha08}). A triangle and a 4-cycle are the smallest possible cycles in a one-mode and bipartite network respectively.  

In a bipartite network a 4-cycle shows that two primary nodes are connected twice via two secondary nodes. However, since the one-mode clustering coefficient measures closure between three nodes, \citet{ops13} chooses to define the clustering coefficient for bipartite networks in terms of paths of length 4 and cycles of length 6:

\begin{equation}\label{eqn:biClust}C^* = \frac{\textrm{closed 4-paths}}{\textrm{4-paths}} = \frac{\tau^*_{\Delta}}{\tau^*},\end{equation} 

where $\tau^*$ is the number of 4-paths and $\tau^*_{\Delta}$ is the number of these 4-paths that are closed. A closed 4-path is equivalent to a cycle of length 6. We give an explanation in Section \ref{sec:formation}. 

The local clustering coefficient of a node $v_i$ is given as follows:

\begin{equation}\label{eqn:localBiClust}C^*(i) = \frac{\tau^*_{i,\Delta}}{\tau^*_i},\end{equation}

where $\tau^*_i$ is the number of 4-paths that are centred at node $v_i$ and $\tau^*_{i,\Delta}$ is the number of these 4-paths that are closed.

A path is defined as a sequence of unique nodes and edges. In other words, when traversing a path in a network, no node is revisited. The length of a path is equal to the number of its edges. A cycle is a path that starts and ends at the same node.  

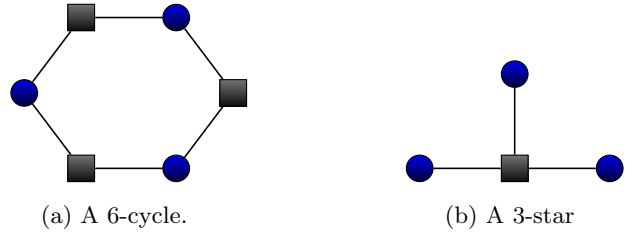
\begin{figure}[h]
	\centering
	\begin{subfigure}[b]{0.22\textwidth}
		\centering
		\scalebox{1.0}{
			\begin{tikzpicture}[node distance=1cm, 
			c/.style={circle, thin, minimum width=1em, minimum height=1em, text centered, top color=blue!90!black, bottom color=blue!20!black, draw=black}, sq/.style={rectangle, thin, 		
				minimum width=1em, minimum height=1em, text centered, top color=gray!90!black, bottom color=gray!10!black, draw=black},
			sl/.style={-,>=stealth',auto,semithick,align = center,draw=black}, dl/.style={dotted,>=stealth',shorten >=1pt,auto,semithick,align = center,draw=black}]
			
			\node[c](c1){};
			\node[c, xshift = 2cm, yshift = 1cm](c2){};
			\node[c, xshift = 2cm, yshift = -1cm](c3){};
			\node[sq, xshift = 0.75cm, yshift = 1cm](r1){};
			\node[sq, xshift = 0.75cm, yshift = -1cm](r2){};
			\node[sq, xshift = 2.75cm, yshift = 0cm](r3){};
			
			\draw[sl] (c1) to node{}(r1);
			\draw[sl] (r1) to node{}(c2);
			\draw[sl] (c2) to node{}(r3);
			\draw[sl] (r3) to node{}(c3);
			\draw[sl] (c3) to node{}(r2);
			\draw[sl] (r2) to node{}(c1);
			
			\end{tikzpicture}
		}
		\caption{A 6-cycle.} \label{im:sixCycle}
	\end{subfigure}
	\hspace{2cm}
	\begin{subfigure}[b]{0.22\textwidth}
		\centering
		\scalebox{1.0}{
			\begin{tikzpicture}[node distance=1cm, 
			c/.style={circle, thin, minimum width=1em, minimum height=1em, text centered, top color=blue!90!black, bottom color=blue!20!black, draw=black}, sq/.style={rectangle, thin, 		
				minimum width=1em, minimum height=1em, text centered, top color=gray!90!black, bottom color=gray!10!black, draw=black},
			sl/.style={-,>=stealth',auto,semithick,align = center,draw=black}, dl/.style={dotted,>=stealth',shorten >=1pt,auto,semithick,align = center,draw=black}]
			
			\node[c](c1){};
			\node[c, xshift = 2.5cm, yshift = 0cm](c2){};
			\node[c, xshift = 1.25cm, yshift = 1.25cm](c3){};
			\node[sq, xshift = 1.25cm, yshift = 0cm](r1){};
			
			\draw[sl] (c1) to node{}(r1);
			\draw[sl] (c2) to node{}(r1);
			\draw[sl] (c3) to node{}(r1);
			\end{tikzpicture}
		}
		\caption{A 3-star}\label{im:threeStar}
	\end{subfigure}
	\caption{There are two bipartite subgraphs that may be considered as a closed connection. In a 6-cycle (a) every primary node is connected to every other primary node via a different secondary node, whereas in a 3-star (b), all three primary nodes are connected to each other via the same secondary node.}\label{im:clustering}
\end{figure}

Since the bipartite clustering coefficient should measure closure between three nodes of the same type (as it does in one-mode networks) the idea of triadic closure is the obvious direction to follow. Figure \ref{im:clustering} shows the two subgraphs that may be considered as a closed connection between three primary nodes. Primary nodes are represented by circles and secondary nodes are represented by squares. However, star subgraphs are the reason for the count of triangles in a projected one-mode network to be higher than expected \citep{ops13}. Hence Eq. \eqref{eqn:biClust}, does not consider this structure as a closed connection between three primary nodes, only counting cycles of length 6.

\section{The structure of bipartite clusters}\label{sec:formation}
This section looks at how clusters in bipartite networks are structured.

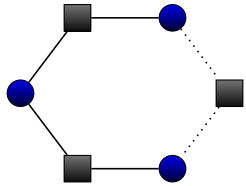
\begin{figure}[h]
	\centering
	\scalebox{1.0}{
		\begin{tikzpicture}[node distance=1cm, 
		c/.style={circle, thin, minimum width=1em, minimum height=1em, text centered, top color=blue!90!black, bottom color=blue!20!black, draw=black}, sq/.style={rectangle, thin, 		
			minimum width=1em, minimum height=1em, text centered, top color=gray!90!black, bottom color=gray!10!black, draw=black},
		sl/.style={-,>=stealth',auto,semithick,align = center,draw=black}, dl/.style={dotted,>=stealth',shorten >=1pt,auto,semithick,align = center,draw=black}]
		
		\node[c](c1){};
		\node[c, xshift = 2cm, yshift = 1cm](c2){};
		\node[c, xshift = 2cm, yshift = -1cm](c3){};
		\node[sq, xshift = 0.75cm, yshift = 1cm](r1){};
		\node[sq, xshift = 0.75cm, yshift = -1cm](r2){};
		\node[sq, xshift = 2.75cm, yshift = 0cm](r3){};
		
		\draw[sl] (c1) to node{}(r1);
		\draw[sl] (r1) to node{}(c2);
		\draw[dl] (c2) to node{}(r3);
		\draw[dl] (r3) to node{}(c3);
		\draw[sl] (c3) to node{}(r2);
		\draw[sl] (r2) to node{}(c1);
		
		\end{tikzpicture}
	}
	\caption{A 6-cycle can be formed by connecting a secondary node to the two end nodes of a 4-path.}\label{im:closed4path}
\end{figure}

A bipartite 6-cycle may be formed by connecting a secondary node to the two end nodes of a 4-path (see Fig. \ref{im:closed4path}). Hence the terminology in Eq. \eqref{eqn:biClust} and Eq. \eqref{eqn:localBiClust}.

\begin{figure}[h]
	\centering
	\begin{subfigure}[t]{0.45\textwidth}
		\centering
		\scalebox{1.0}{
			\begin{tikzpicture}[node distance=1cm, 
			c/.style={circle, thin, minimum width=1em, minimum height=1em, text centered, top color=blue!90!black, bottom color=blue!20!black, draw=black}, sq/.style={rectangle, thin, 		
				minimum width=1em, minimum height=1em, text centered, top color=gray!90!black, bottom color=gray!10!black, draw=black},
			sl/.style={-,>=stealth',auto,semithick,align = center,draw=black}, dl/.style={dotted,>=stealth',shorten >=1pt,auto,semithick,align = center,draw=black}]
			
			\node[c](c1){};
			\node[c, xshift = 2cm, yshift = 1cm](c2){};
			\node[c, xshift = 2cm, yshift = -1cm](c3){};
			\node[sq, xshift = 0.75cm, yshift = 1cm](r1){};
			\node[sq, xshift = 0.75cm, yshift = -1cm](r2){};
			\node[sq, xshift = 2.75cm, yshift = 0cm](r3){};
			
			\draw[sl] (c1) to node{}(r1);
			\draw[sl] (r1) to node{}(c2);
			\draw[sl] (c2) to node{}(r3);
			\draw[sl] (r3) to node{}(c3);
			\draw[sl] (c3) to node{}(r2);
			\draw[sl] (r2) to node{}(c1);
			
			\draw[dl] (c1) to node{}(r3);
			\draw[dl] (c2) to node{}(r2);
			\draw[dl] (c3) to node{}(r1);
			
			\node[c, xshift = 4cm, yshift = 0cm](c1){};
			\node[c, xshift = 6cm, yshift = 1cm](c2){};
			\node[c, xshift = 6cm, yshift = -1cm](c3){};
			\node[sq, xshift = 4.75cm, yshift = -1cm](r2){};
			\node[sq, xshift = 6.75cm, yshift = 0cm](r3){};
			
			\draw[sl] (c2) to node{}(r3);
			\draw[sl] (r3) to node{}(c3);
			\draw[sl] (c3) to node{}(r2);
			\draw[sl] (r2) to node{}(c1);
			
			\draw[dl] (c1) to node{}(r3);
			\draw[dl] (c2) to node{}(r2);
			
			\end{tikzpicture}
		}
		\caption{Possible additional edges (dashed lines) in a bipartite 6-cycle and 4-path. }\label{im:6-cycleIC}
	\end{subfigure}
	
	\vspace{0.5cm}
	\begin{subfigure}[t]{0.45\textwidth}
		\centering
		\scalebox{1.0}{
			\begin{tikzpicture}[node distance=1cm, 
			c/.style={circle, thin, minimum width=1em, minimum height=1em, text centered, top color=blue!90!black, bottom color=blue!20!black, draw=black}, sq/.style={rectangle, thin, 		
				minimum width=1em, minimum height=1em, text centered, top color=gray!90!black, bottom color=gray!10!black, draw=black},
			sl/.style={-,>=stealth',auto,semithick,align = center,draw=black}, dl/.style={dotted,>=stealth',shorten >=1pt,auto,semithick,align = center,draw=black}]
			
			\node[c, xshift = 2.5cm, yshift = -1cm](c1){};
			\node[c, xshift = 0cm, yshift = -1cm](c2){};
			\node[c, xshift = 1.25cm, yshift = 1cm](c3){};
			
			\draw[sl] (c1) to node{}(c2);
			\draw[sl] (c1) to node{}(c3);
			\draw[sl] (c2) to node{}(c3);
			
			\node[c, xshift = 6cm, yshift = -1cm](c1){};
			\node[c, xshift = 3.5cm, yshift = -1cm](c2){};
			\node[c, xshift = 4.75cm, yshift = 1cm](c3){};
			
			\draw[sl] (c1) to node{}(c2);
			\draw[sl] (c2) to node{}(c3);
			
			\end{tikzpicture}
		}
		\caption{It is impossible to connect the nodes of a triangle and 2-path with additional edges without introducing additional edges. }\label{im:4-pathsIc}
	\end{subfigure}
	\caption{{\bf(a)} A bipartite cycle of length 6 can have at most three additional edges, whereas a bipartite 4-path can have at most two. Additional edges are represented by dashed lines. {\bf(b)} In a one-mode network, no additional edges can be added to a triangle or a 2-path without creating multiple edges.}\label{im:innerConnections}
\end{figure}
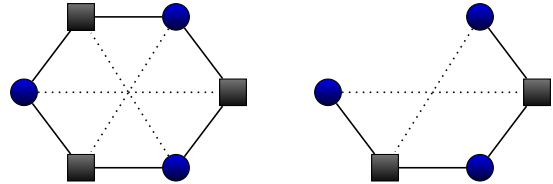
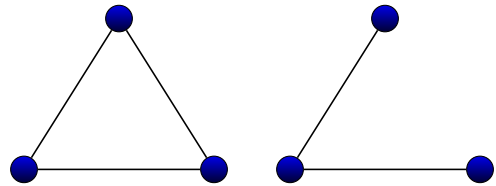

There is an important difference in the cluster formation in bipartite networks as opposed to the formation in one-mode networks. In a one-mode network, additional edges between the nodes of a triangle and between the nodes of a 2-path cannot exist without introducing multiple edges. However, in a bipartite network, the nodes of a 6-cycle as well as the nodes of a 4-path may be connected to each other by additional edges (see Fig. \ref{im:innerConnections}).  Additional edges in the bipartite subgraphs give rise to different structures with distinct meanings that depend on the network in question.

\begin{figure}[h]
	\centering
	\begin{subfigure}[t]{0.22\textwidth}
		\scalebox{1.0}{
			\begin{tikzpicture}[node distance=1cm, 
			c/.style={circle, thin, minimum width=1em, minimum height=1em, text centered, top color=blue!90!black, bottom color=blue!20!black, draw=black}, sq/.style={rectangle, thin, 		
				minimum width=1em, minimum height=1em, text centered, top color=gray!90!black, bottom color=gray!10!black, draw=black},
			sl/.style={-,>=stealth',auto,semithick,align = center,draw=black}, dl/.style={dotted,>=stealth',shorten >=1pt,auto,semithick,align = center,draw=black}]
			
			\node[c](c1){};
			\node[c, xshift = 2cm, yshift = 1cm](c2){};
			\node[c, xshift = 2cm, yshift = -1cm](c3){};
			\node[sq, xshift = 0.75cm, yshift = 1cm](r1){};
			\node[sq, xshift = 0.75cm, yshift = -1cm](r2){};
			\node[sq, xshift = 2.75cm, yshift = 0cm](r3){};
			
			\draw[sl] (c1) to node{}(r1);
			\draw[sl] (r1) to node{}(c2);
			\draw[sl] (c2) to node{}(r3);
			\draw[sl] (r3) to node{}(c3);
			\draw[sl] (c3) to node{}(r2);
			\draw[sl] (r2) to node{}(c1);
			\end{tikzpicture}
		}
		\caption{An unconnected 6-cycle. }\label{im:noIC6cycle}
	\end{subfigure}
	\hspace{2cm}
	\begin{subfigure}[t]{0.22\textwidth}
		\scalebox{1.0}{
			\begin{tikzpicture}[node distance=1cm, 
			c/.style={circle, thin, minimum width=1em, minimum height=1em, text centered, top color=blue!90!black, bottom color=blue!20!black, draw=black}, sq/.style={rectangle, thin, 		
				minimum width=1em, minimum height=1em, text centered, top color=gray!90!black, bottom color=gray!10!black, draw=black},
			sl/.style={-,>=stealth',auto,semithick,align = center,draw=black}, dl/.style={dotted,>=stealth',shorten >=1pt,auto,semithick,align = center,draw=black}]
			
			\node[c](c1){};
			\node[c, xshift = 2cm, yshift = 1cm](c2){};
			\node[c, xshift = 2cm, yshift = -1cm](c3){};
			\node[sq, xshift = 0.75cm, yshift = 1cm](r1){};
			\node[sq, xshift = 0.75cm, yshift = -1cm](r2){};
			\node[sq, xshift = 2.75cm, yshift = 0cm](r3){};
			
			\draw[sl] (c1) to node{}(r1);
			\draw[sl] (r1) to node{}(c2);
			\draw[sl] (c2) to node{}(r3);
			\draw[sl] (r3) to node{}(c3);
			\draw[sl] (c3) to node{}(r2);
			\draw[sl] (r2) to node{}(c1);
			
			\draw[dl] (r2) to node{}(c2);
			\end{tikzpicture}
		}
		\caption{A sparsely connected 6-cycle. }\label{im:oneIC6cycle}
	\end{subfigure}
	
	\begin{subfigure}[t]{0.22\textwidth}
		\scalebox{1.0}{
			\begin{tikzpicture}[node distance=1cm, 
			c/.style={circle, thin, minimum width=1em, minimum height=1em, text centered, top color=blue!90!black, bottom color=blue!20!black, draw=black}, sq/.style={rectangle, thin, 		
				minimum width=1em, minimum height=1em, text centered, top color=gray!90!black, bottom color=gray!10!black, draw=black},
			sl/.style={-,>=stealth',auto,semithick,align = center,draw=black}, dl/.style={dotted,>=stealth',shorten >=1pt,auto,semithick,align = center,draw=black}]
			
			\node[c](c1){};
			\node[c, xshift = 2cm, yshift = 1cm](c2){};
			\node[c, xshift = 2cm, yshift = -1cm](c3){};
			\node[sq, xshift = 0.75cm, yshift = 1cm](r1){};
			\node[sq, xshift = 0.75cm, yshift = -1cm](r2){};
			\node[sq, xshift = 2.75cm, yshift = 0cm](r3){};
			
			\draw[sl] (c1) to node{}(r1);
			\draw[sl] (r1) to node{}(c2);
			\draw[sl] (c2) to node{}(r3);
			\draw[sl] (r3) to node{}(c3);
			\draw[sl] (c3) to node{}(r2);
			\draw[sl] (r2) to node{}(c1);	
			
			\draw[dl] (r2) to node{}(c2);
			\draw[dl] (r1) to node{}(c3);
			\end{tikzpicture}
		}
		\caption{A highly connected 6-cycle.} \label{im:twoIC6cycle}
	\end{subfigure}
	\hspace{2cm}
	\begin{subfigure}[t]{0.22\textwidth}
		\scalebox{1.0}{
			\begin{tikzpicture}[node distance=1cm, 
			c/.style={circle, thin, minimum width=1em, minimum height=1em, text centered, top color=blue!90!black, bottom color=blue!20!black, draw=black}, sq/.style={rectangle, thin, 		
				minimum width=1em, minimum height=1em, text centered, top color=gray!90!black, bottom color=gray!10!black, draw=black},
			sl/.style={-,>=stealth',auto,semithick,align = center,draw=black}, dl/.style={dotted,>=stealth',shorten >=1pt,auto,semithick,align = center,draw=black}]
			
			\node[c](c1){};
			\node[c, xshift = 2cm, yshift = 1cm](c2){};
			\node[c, xshift = 2cm, yshift = -1cm](c3){};
			\node[sq, xshift = 0.75cm, yshift = 1cm](r1){};
			\node[sq, xshift = 0.75cm, yshift = -1cm](r2){};
			\node[sq, xshift = 2.75cm, yshift = 0cm](r3){};	
			
			\draw[sl] (c1) to node{}(r1);
			\draw[sl] (r1) to node{}(c2);
			\draw[sl] (c2) to node{}(r3);
			\draw[sl] (r3) to node{}(c3);
			\draw[sl] (c3) to node{}(r2);
			\draw[sl] (r2) to node{}(c1);	
			
			\draw[dl] (c1) to node{}(r3);
			\draw[dl] (r2) to node{}(c2);
			\draw[dl] (r1) to node{}(c3);
			\end{tikzpicture}
		}
		\caption{A completely connected 6-cycle.} \label{im:threeIC6cycle}
	\end{subfigure}
	\caption{All possible structures that a bipartite 6-cycle may have.}\label{im:innerConnections6cycle}
\end{figure}
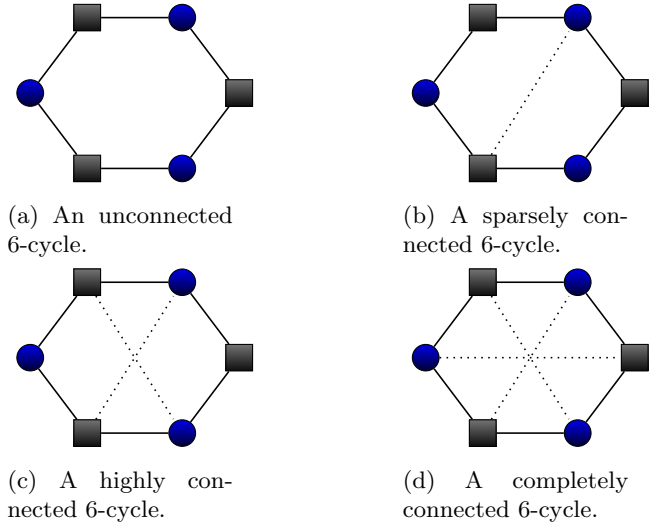

It is possible to form different 6-cycles between the same six nodes by traversing different edges. The number of different 6-cycles between the same set of nodes depends on the number of edges that connect these six nodes to each other. A bipartite 6-cycle can have at most three additional edges (see Fig. \ref{im:innerConnections6cycle}). We call the structures shown in Fig. \ref{im:noIC6cycle}, \ref{im:oneIC6cycle}, \ref{im:twoIC6cycle} and \ref{im:threeIC6cycle} an unconnected 6-cycle, a sparsely connected 6-cycle, a highly connected 6-cycle and a completely connected 6-cycle respectively.

By traversing the different edges in the structures shown in Fig. \ref{im:innerConnections6cycle}, one can confirm that an unconnected 6-cycle contributes one to the overall count of 6-cycles. A sparsely connected 6-cycle contains a single 6-cycle and hence also contributes one to the overall count of 6-cycles. A highly connected 6-cycle contributes two to the overall count of 6-cycles and finally, a completely connected 6-cycle contributes six to the overall count of 6-cycles. As we look at undirected networks the direction of the cycle does not matter. It is also irrelevant at which of the six nodes the cycle starts. 

This clearly shows that counting the number of 6-cycles and not distinguishing the structures shown in Fig. \ref{im:innerConnections6cycle} leads to an over count of 6-cycles. Our clustering coefficient treats each of the different types of 6-cycles separately. Distinguishing between the different structures gives insight into how interconnected any set of three nodes of the same type can be. 

\section{A new improved bipartite clustering coefficient}\label{sec:improvedClustering}
We now give a bipartite clustering coefficient that distinguishes between the 6-cycles identified in Section \ref{sec:formation}. We develop four bipartite clustering coefficients, one for each type of 6-cycle, that provide new information about the network not revealed by previously defined clustering coefficients.

In order to calculate the clustering coefficient, we need to determine all possibilities by which the different types of 6-cycles may be formed. We wish to analyse a snapshot of a network at a particular point in time, however, this needs awareness of how the network was formed over time. But first we show why it is necessary to distinguish between different types of bipartite networks. 

\begin{figure}[h]
	\centering
	\scalebox{0.8}{
		\begin{tikzpicture}[node distance=1cm, 
		c/.style={circle, thin, minimum width=1em, minimum height=1em, text centered, top color=blue!90!black, bottom color=blue!20!black, draw=black}, sq/.style={rectangle, thin, 		
			minimum width=1em, minimum height=1em, text centered, top color=gray!90!black, bottom color=gray!10!black, draw=black},
		sl/.style={-,>=stealth',auto,semithick,align = center,draw=black}, dl/.style={dotted,>=stealth',shorten >=1pt,auto,semithick,align = center,draw=black}]
		
		\node[c, label=left:{person a}](c1){};
		\node[c, label=right:{person b}, xshift = 3cm, yshift = 1.5cm](c2){};
		\node[c, label=right:{person c}, xshift = 3cm, yshift = -1.5cm](c3){};
		\node[sq, label=above:{event 1, $t_1$}, xshift = 1.25cm, yshift = 1.5cm](r1){};
		\node[sq, label=below:{event 2, $t_2$}, xshift = 1.25cm, yshift = -1.5cm](r2){};
		\node[sq, label=right:{event 3, $t_3$}, xshift = 4.25cm, yshift = 0cm](r3){};
		
		\draw[sl] (c1) to node{}(r1);
		\draw[sl] (r1) to node{}(c2);
		\draw[sl] (c2) to node{}(r3);
		\draw[sl] (r3) to node{}(c3);
		\draw[sl] (c3) to node{}(r2);
		\draw[sl] (r2) to node{}(c1);
		\end{tikzpicture}
	}
	\caption{Person $a$ and person $b$ attended event 1 that took place at time $t_1$. Person $a$ and person $c$ attended event 2 that took place at time $t_2$. Finally, person $b$ and person $c$ attended event 3 that took place at time $t_3$. At time $t_3$ the 6-cycle is complete and no connections within the cycle can be formed at a later point in time.}\label{im:exampleType1}
\end{figure}
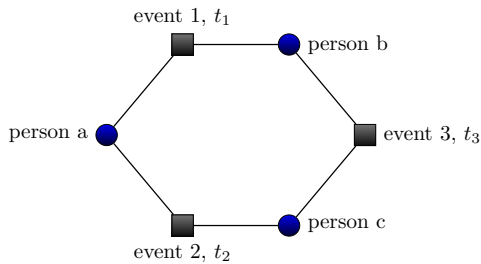

In some bipartite networks a primary node can only connect to a particular secondary node at a particular point in time. We call these networks time dependent networks. In a network that models the attendance of people at events, where each event takes place at a specific point in time, ties cannot be formed within an existing 6-cycle. For instance, assume there exists a 6-cycle that is part of a network of people and events (see Fig. \ref{im:exampleType1}). Events 1, 2 and 3 take place at times $t_1, t_2$ and $t_3$ respectively. Figure \ref{im:exampleType1} clearly shows that connections within 6-cycles cannot be formed at a later point in time, as all three events have passed.  

In networks where nodes are allowed to connect to secondary nodes at any point in time, it is possible to form connections within an existing 6-cycle. A network of online forums and users, where the forums form the secondary node set, is an example of such a network. Each forum is accessible over a period of time and hence it is possible for connections to be formed within an existing 6-cycle. It is obvious that the origins of the 6-cycles shown in Fig. \ref{im:innerConnections6cycle} are different in the two types of networks. Here, we only consider time dependent networks, as defined in the paragraph above. 

\begin{figure}[h]
	\centering
	\begin{subfigure}[b]{0.3\textwidth}
		\centering
		\scalebox{0.45}{
			\begin{tikzpicture}[node distance=1cm, 
			c/.style={circle, thin, minimum width=1em, minimum height=1em, text centered, top color=blue!90!black, bottom color=blue!20!black, draw=black}, sq/.style={rectangle, thin, 		
				minimum width=1em, minimum height=1em, text centered, top color=gray!90!black, bottom color=gray!10!black, draw=black},
			sl/.style={-,>=stealth',auto,semithick,align = center,draw=black}, dl/.style={dotted,>=stealth',shorten >=1pt,auto,semithick,align = center,draw=black}]
			
			\node[c, label=left:{$v_0$}](c1){};
			\node[c, label=right:{$v_1$}, xshift = 2cm, yshift = 1cm](c2){};
			\node[c, label=right:{$v_2$}, xshift = 2cm, yshift = -1cm](c3){};
			\node[sq, label=left:{$w_0$}, xshift = 0.75cm, yshift = 1cm](r1){};
			\node[sq, label=left:{$w_2$},xshift = 0.75cm, yshift = -1cm](r2){};
			\node[sq, label=right:{$w_1$}, xshift = 2.75cm, yshift = 0cm](r3){};	
			
			\draw[sl] (c1) to node{}(r1);
			\draw[sl] (r1) to node{}(c2);
			\draw[sl] (c2) to node{}(r3);
			\draw[sl] (r3) to node{}(c3);
			\draw[sl] (c3) to node{}(r2);
			\draw[sl] (r2) to node{}(c1);

			\node[c, label=left:{$v_0$}, xshift = 4.75cm, yshift = 0cm](c4){};
			\node[c, label=right:{$v_1$}, xshift = 6.75cm, yshift = 1cm](c5){};
			\node[c, label=right:{$v_2$}, xshift = 6.75cm, yshift = -1cm](c6){};
			\node[sq, label=left:{$w_0$}, xshift = 5.5cm, yshift = 1cm](r4){};
			\node[sq, label=left:{$w_2$},xshift = 5.5cm, yshift = -1cm](r5){};
			\node[sq, label=right:{$w_1$}, xshift = 7.5cm, yshift = 0cm](r6){};
			
			\draw[sl] (c4) to node{}(r4);
			\draw[sl] (r4) to node{}(c5);
			\draw[sl] (c5) to node{}(r6);
			\draw[sl] (r6) to node{}(c6);
			\draw[sl] (c6) to node{}(r5);
			\draw[sl] (r5) to node{}(c4);	
			
			\draw[dl] (c5) to node{}(r5);
			
			\node[c, label=left:{$v_0$}, xshift = 1.25cm, yshift = 4cm](c7){};
			\node[c, label=above:{$v_1$}, xshift = 3.75cm, yshift = 5.5cm](c8){};
			\node[c, label=right:{$v_2$}, xshift = 6.25cm, yshift = 4cm](c9){};
			\node[sq, label=left:{$w_0$}, xshift = 2.5cm, yshift = 5cm](r7){};
			\node[sq, label=right:{$w_1$}, xshift = 5cm, yshift = 5cm](r8){};
			
			\draw[sl] (c7) to node{}(r7);
			\draw[sl] (r7) to node{}(c8);
			\draw[sl] (c8) to node{}(r8);
			\draw[sl] (r8) to node{}(c9);
			
			\draw[thick,->] (1.5,3.5) -- (1.5,1.5);
			\draw[thick,->] (6,3.5) -- (6,1.5);
			\end{tikzpicture}
		}
		\caption{}\label{im:4-pathNoIC}
	\end{subfigure}
	\quad
	\begin{subfigure}[b]{0.3\textwidth}
		\centering
		\scalebox{0.45}{
			\begin{tikzpicture}[node distance=1cm, 
			c/.style={circle, thin, minimum width=1em, minimum height=1em, text centered, top color=blue!90!black, bottom color=blue!20!black, draw=black}, sq/.style={rectangle, thin, 		
				minimum width=1em, minimum height=1em, text centered, top color=gray!90!black, bottom color=gray!10!black, draw=black},
			sl/.style={-,>=stealth',auto,semithick,align = center,draw=black}, dl/.style={dotted,>=stealth',shorten >=1pt,auto,semithick,align = center,draw=black}]
			
			\node[c, label=left:{$v_0$}](c1){};
			\node[c, label=right:{$v_1$}, xshift = 2cm, yshift = 1cm](c2){};
			\node[c, label=right:{$v_2$}, xshift = 2cm, yshift = -1cm](c3){};
			\node[sq, label=left:{$w_0$}, xshift = 0.75cm, yshift = 1cm](r1){};
			\node[sq, label=left:{$w_2$},xshift = 0.75cm, yshift = -1cm](r2){};
			\node[sq, label=right:{$w_1$}, xshift = 2.75cm, yshift = 0cm](r3){};	
			
			\draw[sl] (c1) to node{}(r1);
			\draw[sl] (r1) to node{}(c2);
			\draw[sl] (c2) to node{}(r3);
			\draw[sl] (r3) to node{}(c3);
			\draw[sl] (c3) to node{}(r2);
			\draw[sl] (r2) to node{}(c1);
			
			\draw[dl] (c1) to node{}(r3);
			
			\node[c, label=left:{$v_0$}, xshift = 4.75cm, yshift = 0cm](c4){};
			\node[c, label=right:{$v_1$}, xshift = 6.75cm, yshift = 1cm](c5){};
			\node[c, label=right:{$v_2$}, xshift = 6.75cm, yshift = -1cm](c6){};
			\node[sq, label=left:{$w_0$}, xshift = 5.5cm, yshift = 1cm](r4){};
			\node[sq, label=left:{$w_2$},xshift = 5.5cm, yshift = -1cm](r5){};
			\node[sq, label=right:{$w_1$}, xshift = 7.5cm, yshift = 0cm](r6){};
			
			\draw[sl] (c4) to node{}(r4);
			\draw[sl] (r4) to node{}(c5);
			\draw[sl] (c5) to node{}(r6);
			\draw[sl] (r6) to node{}(c6);
			\draw[sl] (c6) to node{}(r5);
			\draw[sl] (r5) to node{}(c4);	
			
			\draw[dl] (c4) to node{}(r6);
			\draw[dl] (c5) to node{}(r5);
			
			\node[c, label=left:{$v_0$}, xshift = 1.25cm, yshift = 4cm](c7){};
			\node[c, label=above:{$v_1$}, xshift = 3.75cm, yshift = 5.5cm](c8){};
			\node[c, label=right:{$v_2$}, xshift = 6.25cm, yshift = 4cm](c9){};
			\node[sq, label=left:{$w_0$}, xshift = 2.5cm, yshift = 5cm](r7){};
			\node[sq, label=right:{$w_1$}, xshift = 5cm, yshift = 5cm](r8){};
			
			\draw[sl] (c7) to node{}(r7);
			\draw[sl] (r7) to node{}(c8);
			\draw[sl] (c8) to node{}(r8);
			\draw[sl] (r8) to node{}(c9);
			
			\draw[dl] (c7) to node{}(r8);
			
			\draw[thick,->] (1.5,3.5) -- (1.5,1.5);
			\draw[thick,->] (6,3.5) -- (6,1.5);
			\end{tikzpicture}
		}
		\caption{}\label{im:4-pathOneIC}
	\end{subfigure}
	\quad
	\begin{subfigure}[b]{0.3\textwidth}
		\centering
		\scalebox{0.45}{
			\begin{tikzpicture}[node distance=1cm, 
			c/.style={circle, thin, minimum width=1em, minimum height=1em, text centered, top color=blue!90!black, bottom color=blue!20!black, draw=black}, sq/.style={rectangle, thin, 		
				minimum width=1em, minimum height=1em, text centered, top color=gray!90!black, bottom color=gray!10!black, draw=black},
			sl/.style={-,>=stealth',auto,semithick,align = center,draw=black}, dl/.style={dotted,>=stealth',shorten >=1pt,auto,semithick,align = center,draw=black}]
			
			\node[c, label=left:{$v_0$}](c1){};
			\node[c, label=right:{$v_1$}, xshift = 2cm, yshift = 1cm](c2){};
			\node[c, label=right:{$v_2$}, xshift = 2cm, yshift = -1cm](c3){};
			\node[sq, label=left:{$w_0$}, xshift = 0.75cm, yshift = 1cm](r1){};
			\node[sq, label=left:{$w_2$},xshift = 0.75cm, yshift = -1cm](r2){};
			\node[sq, label=right:{$w_1$}, xshift = 2.75cm, yshift = 0cm](r3){};	
			
			\draw[sl] (c1) to node{}(r1);
			\draw[sl] (r1) to node{}(c2);
			\draw[sl] (c2) to node{}(r3);
			\draw[sl] (r3) to node{}(c3);
			\draw[sl] (c3) to node{}(r2);
			\draw[sl] (r2) to node{}(c1);
			
			\draw[dl] (c1) to node{}(r3);
			\draw[dl] (c3) to node{}(r1);
			
			\node[c, label=left:{$v_0$}, xshift = 4.75cm, yshift = 0cm](c4){};
			\node[c, label=right:{$v_1$}, xshift = 6.75cm, yshift = 1cm](c5){};
			\node[c, label=right:{$v_2$}, xshift = 6.75cm, yshift = -1cm](c6){};
			\node[sq, label=left:{$w_0$}, xshift = 5.5cm, yshift = 1cm](r4){};
			\node[sq, label=left:{$w_2$},xshift = 5.5cm, yshift = -1cm](r5){};
			\node[sq, label=right:{$w_1$}, xshift = 7.5cm, yshift = 0cm](r6){};
			
			\draw[sl] (c4) to node{}(r4);
			\draw[sl] (r4) to node{}(c5);
			\draw[sl] (c5) to node{}(r6);
			\draw[sl] (r6) to node{}(c6);
			\draw[sl] (c6) to node{}(r5);
			\draw[sl] (r5) to node{}(c4);	
			
			\draw[dl] (c4) to node{}(r6);
			\draw[dl] (c5) to node{}(r5);
			\draw[dl] (c6) to node{}(r4);
			
			\node[c, label=left:{$v_0$}, xshift = 1.25cm, yshift = 4cm](c7){};
			\node[c, label=above:{$v_1$}, xshift = 3.75cm, yshift = 5.5cm](c8){};
			\node[c, label=right:{$v_2$}, xshift = 6.25cm, yshift = 4cm](c9){};
			\node[sq, label=left:{$w_0$}, xshift = 2.5cm, yshift = 5cm](r7){};
			\node[sq, label=right:{$w_1$}, xshift = 5cm, yshift = 5cm](r8){};
			
			\draw[sl] (c7) to node{}(r7);
			\draw[sl] (r7) to node{}(c8);
			\draw[sl] (c8) to node{}(r8);
			\draw[sl] (r8) to node{}(c9);
			
			\draw[dl] (c7) to node{}(r8);
			\draw[dl] (c9) to node{}(r7);
			
			\draw[thick,->] (1.5,3.5) -- (1.5,1.5);
			\draw[thick,->] (6,3.5) -- (6,1.5);
			\end{tikzpicture}
		}
		\caption{}\label{im:4-pathTwoIC}
	\end{subfigure}
	\caption{All possibilities by which the different structured 6-cycles can be formed in a time dependent network. An unconnected 6-cycle and a completely connected 6-cycle can each only originate from one, distinct 4-path. A sparsely connected and a highly connected 6-cycle each have two origins.}\label{im:formation6cycles}
\end{figure}
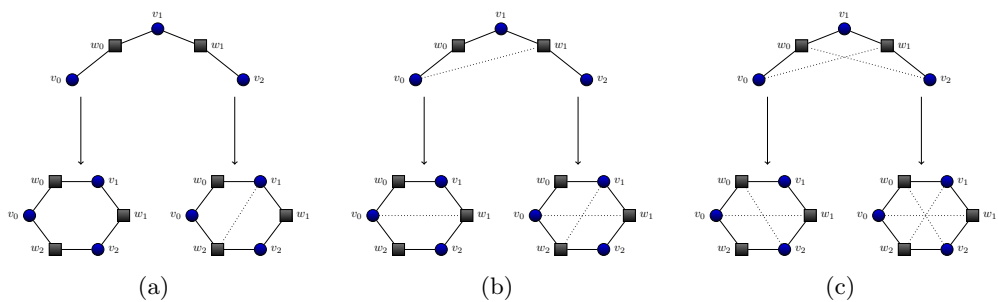

Any 6-cycle in a time dependent bipartite network is formed by connecting a secondary node to the two primary end nodes of a 4-path. Figure \ref{im:formation6cycles} shows all the possibilities by which the distinct 6-cycles may be formed. We call the 4-path in Fig. \ref{im:4-pathNoIC} an unconnected 4-path, the 4-path in Fig. \ref{im:4-pathOneIC} a connected 4-path and the 4-path in Fig. \ref{im:4-pathTwoIC} a completely connected 4-path. 

Using the origins of 6-cycles, we define equations \eqref{eq:cc0} - \eqref{eq:cc3} to measure four different clustering coefficients $cc_{(k)}$ in a time dependent bipartite network.

The unconnected clustering coefficient:

\begin{equation}\label{eq:cc0}
cc_{(0)} = \frac{\lambda_{(0)}^*}{\lambda_{(0)}},
\end{equation}

where $\lambda^*_{(0)}$ is the number of closed 4-paths that form an unconnected 6-cycle and $\lambda_{(0)}$ is the total number of unconnected 4-paths. The unconnected clustering coefficient $cc_{(0)}$ measures the proportion of unconnected 4-paths that are closed and form an unconnected 6-cycle.  

The sparsely connected clustering coefficient:

\begin{equation}\label{eq:cc1}
cc_{(1)} = \frac{\lambda_{(1)}^* }{\lambda_{(0)} + \lambda_{(1)}},
\end{equation}

where $\lambda^*_{(1)}$ is the number of closed 4-paths that form a sparsely connected 6-cycle and $\lambda_{(1)}$ is the total number of connected 4-paths. The sparsely connected clustering coefficient $cc_{(1)}$ measures the proportion of 4-paths that are closed and form a sparsely connected 6-cycle.  

The highly connected clustering coefficient:

\begin{equation}\label{eq:cc2}
cc_{(2)} = \frac{\lambda_{(2)}^* }{\lambda_{(1)} + \lambda_{(2)}},
\end{equation}

where $\lambda^*_{(2)}$ is the number of closed 4-paths that form a highly connected 6-cycle and $\lambda_{(2)}$ is the total number of completely connected 4-paths. The highly connected clustering coefficient $cc_{(2)}$ measures the proportion of 4-paths that are closed and form a highly connected 6-cycle.  

The completely connected clustering coefficient:

\begin{equation}\label{eq:cc3}
cc_{(3)} = \frac{\lambda_{(3)}^*}{\lambda_{(2)}},
\end{equation}

where $\lambda^*_{(3)}$ is the number of closed 4-paths that form a completely connected 6-cycle. The clustering coefficient $cc_{(3)}$ measures the proportion of 4-paths that are closed and form a completely connected 6-cycle.  

The local clustering coefficients $cc_{(i,k)}$ of a node $v_i$ can be measured in a similar manner. For example, the local clustering coefficient $cc_{(i,0)}$ of the node $v_i$ is measured by dividing the number of closed 4-paths that are centred at $v_i$ and form an unconnected 6-cycle by the number of all unconnected 4-paths that are centred at $v_i$.

\section{Identifying Important Nodes}\label{sec:importantNodes}

In order to find the driving nodes of a network, we calculate a score that indicates the extent to which a node is driving the clustering behaviour of the complete network, by first comparing the global clustering coefficients to the clustering coefficients of random networks and then comparing the local clustering coefficients of each node to the global clustering coefficients. 

There are two cases: 

\begin{enumerate}
	\item $cc_{(k)} < CI_{(k)}$ or
	\item $cc_{(k)} \geq CI_{(k)}$,
\end{enumerate}

where $CI_{(k)}$ is the mid point of the confidence interval calculated for $cc_{(k)}$ in an ensemble of random networks. The random networks need to have the same size, density and degree distribution as the original network. At least 100 random networks have to be generated in order to to achieve a small enough confidence interval. To be able to compare the scores of different nodes, we first calculate a global driving score, denoted $ds_{(global)}$, for the network by measuring how far each of the global clustering coefficients $cc_{(k)}$ for the given network is from $CI_{(k)}$ and then averaging over all four scores, see Eq. \eqref{gds}. Note that $ds_{(global)}$ lies between 0 and 1. The greater the difference between the global clustering coefficients of the given network $cc_{(k)}$ and the respective $CI_{(k)}$, the higher the global driving score.  

\begin{equation}\label{gds}
ds_{(global)} = \frac{1}{4}\sum_{k=0}^{3} g(k),
\end{equation}

where
\begin{equation}\label{gk}
g(k) = \left\{
\begin{array}{ll}
\frac{|CI_{(k)} - cc_{(k)}|}{CI_{(k)}} & \textrm{if}\quad cc_{(k)} <  CI_{(k)}, \\
\frac{|CI_{(k)} - cc_{(k)}|}{1 - CI_{(k)}} & \textrm{if}\quad cc_{(k)} \geq  CI_{(k)}.
\end{array}
\right.
\end{equation}

A node with a local clustering coefficient $cc_{(i,k)}$ that is close to $CI_{(k)}$, behaves as expected and hence does not contribute to a clustering behaviour that is different to a random network. If the global clustering coefficient $cc_{(k)}$ of the given network is smaller than $CI_{(k)}$, i.e. $cc_{(k)} < CI_{(k)}$, then either

\begin{enumerate}
	\item $cc_{(i,k)} < CI_{(k)}$ or 
	\item $cc_{(i,k)} \geq CI_{(k)}$.
\end{enumerate}

If the local clustering coefficient $cc_{(i,k)}$ of node $i$ also lies below $CI_{(k)}$, then node $i$ contributes to the global clustering behaviour of the whole network and we assign a score between 0 and 1 to node $i$, depending on the difference between the local clustering coefficient and $CI_{(k)}$. If on the other hand, $cc_{(i,k)}$ lies above $CI_{(k)}$ then node $i$ drives against the clustering behaviour and we assign node $i$ a score between 0 and -1. Similarly, when the global clustering coefficient $cc_{(k)}$ lies above the mid point of the confidence interval, there are two cases.  

The driving score, $ds_{(i)}$, of node $i$ is thus given by the following equation:

\begin{equation}\label{ds}
ds_{(i)} = \frac{1}{4}\sum_{k=0}^{3} f(k),
\end{equation}

where

\begin{equation}\label{fk}
f(k) = \left\{
\begin{array}{ll}
\frac{|CI_{(k)} - cc_{(i,k)}|}{CI_{(k)}} & \textrm{if}\quad cc_{(k)} <  CI_{(k)} > cc_{(i,k)}, \\
-\frac{|CI_{(k)} - cc_{(i,k)}|}{1 - CI_{(k)}} & \textrm{if}\quad cc_{(k)} <  CI_{(k)} \leq cc_{(i,k)}, \\
\frac{|CI_{(k)} - cc_{(i,k)}|}{1 - CI_{(k)}} & \textrm{if}\quad cc_{(k)} \geq  CI_{(k)} \leq cc_{(i,k)},\\
-\frac{|CI_{(k)} - cc_{(i,k)}|}{CI_{(k)}} & \textrm{if}\quad cc_{(k)} \geq  CI_{(k)} > cc_{(i,k)}.
\end{array}
\right.
\end{equation}

The four different structures, shown in Fig. \ref{im:innerConnections6cycle}, are considered to be equally important, hence the factor of $\frac{1}{4}$ in Eq. \eqref{ds}. The driving score depends solely on how the network under investigation compares to random networks. 

\begin{figure}[h]
	\centering
	\scalebox{0.8}{
		\begin{tikzpicture}[node distance=1cm, 
		c/.style={circle, thin, minimum width=1em, minimum height=1em, text centered, top color=blue!90!black, bottom color=blue!20!black, draw=black}, sq/.style={rectangle, thin, 		
			minimum width=1em, minimum height=1em, text centered, top color=gray!90!black, bottom color=gray!10!black, draw=black},
		sl/.style={-,>=stealth',auto,semithick,align = center,draw=black}, dl/.style={dotted,>=stealth',shorten >=1pt,auto,semithick,align = center,draw=black}]
		
		\node[] at (1.5,3) {Case I: $cc_{(k)} < CI_{(k)}$};
		\node[] at (0,0) {$ds_{(i)}$};
		\node[] at (0.8,0) {0};
		\node[] at (0.8,1.9) {1};
		\node[] at (0.8,-1.9) {-1};
		\node[] at (3,1.9) {$CI_{(k)} \gg cc_{(i,k)}$};
		\node[] at (3,0) {$CI_{(k)} = cc_{(i,k)}$};
		\node[] at (3,-1.9) {$CI_{(k)} \ll cc_{(i,k)}$};

		\draw[thick,-] (1,2) -- (1,-2);
		\draw[thick,-] (0.9,0) -- (1,0);
		\draw[thick,-] (0.9,1.9) -- (1,1.9);
		\draw[thick,-] (0.9,-1.9) -- (1,-1.9);
		
		\node[] at (7.5,3) {Case II: $cc_{(k)} \geq CI_{(k)}$};
		\node[] at (6,0) {$ds_{(i)}$};
		\node[] at (6.8,0) {0};
		\node[] at (6.8,1.9) {1};
		\node[] at (6.8,-1.9) {-1};
		\node[] at (9,1.9) {$CI_{(k)} \ll cc_{(i,k)}$};
		\node[] at (9,0) {$CI_{(k)} = cc_{(i,k)}$};
		\node[] at (9,-1.9) {$CI_{(k)} \gg cc_{(i,k)}$};
		
		\draw[thick,-] (7,2) -- (7,-2);
		\draw[thick,-] (6.9,0) -- (7,0);
		\draw[thick,-] (6.9,1.9) -- (7,1.9);
		\draw[thick,-] (6.9,-1.9) -- (7,-1.9);

		\end{tikzpicture}
	}
	\caption{If $cc_{(k)} < CI_{(k)}$ node $i$ can only get a high driving score if $CI_{(k)} \gg cc_{(i,k)}$. Similarly, if  $cc_{(k)} \geq CI_{(k)}$ node $i$ can only get a high driving score if $CI_{(k)} \ll cc_{(i,k)}$}\label{im:ds}
\end{figure}
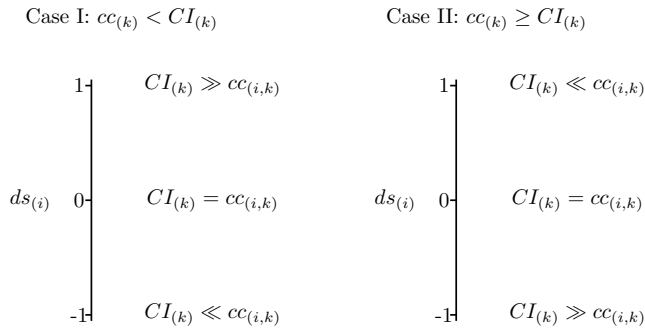

In order to achieve a high driving score, a node does not necessarily have to have high clustering coefficients (see Fig. \ref{im:ds}). For instance, if the global clustering coefficients are low, a node that also has low clustering coefficients, receives a high driving score.

In the following sections we apply the four clustering coefficients (Eq. \eqref{eq:cc0} - Eq. \eqref{eq:cc3}) to different networks from distinct areas and analyse the obtained results. We further identify the important nodes in the different networks.
 
\section{The Southern Women Network}\label{sec:women}
We first calculate the clustering coefficients of a popular, often analysed, data set that was collected by \citet{dav41}. This so called southern women network consists of 18 women and 14 events. An edge between a woman and an  event only exists if the woman attended the event. This bipartite network is clearly a time dependent network as the events take place at a certain time only and cannot be attended afterwards. 

Table \ref{resultsWomen} shows the clustering coefficients of the southern women network and the coefficients of 100 randomly generated bipartite networks and their 95\% confidence intervals. 

\begin{table}[h]
\caption{The four clustering coefficients of the southern women network and the average clustering coefficients of 100 randomly generated networks with their 95\% confidence intervals. }\label{resultsWomen}
	\scriptsize
	\centering
	\begin{tabular}{lll}
		\toprule
		& Southern Women network & random networks\\
		\midrule
		$cc_{(0)}$ & 0.4446 & [0.6261, 0.6478]\\
		$cc_{(1)}$ & 0.6532 & [0.5483, 0.5658]\\
		$cc_{(2)}$ & 0.5984 & [0.3972, 0.4237]\\
		$cc_{(3)}$ & 0.5604 & [0.3018, 0.3457] \\
		\bottomrule
	\end{tabular}
\end{table}

We found that none of the clustering coefficients lie within the 95\% confidence intervals and hence, none of the values are as expected in a random network. The coefficient $cc_{(0)}$ lies below the lower bound of the confidence interval whereas $cc_{(1)}, cc_{(2)}$ and $cc_{(3)}$ lie above the interval. In the average random network $cc_{(0)}$ has the highest value ($cc_{(0)} = 0.6226$), as opposed to the southern women network, where $cc_{(0)}$ takes the lowest value ($cc_{(0)}=0.4446$). Hence, a greater proportion of 4-paths are closed to form an unconnected 6-cycle in random networks than in the southern women network. The remaining three clustering coefficients lie above the 95\% confidence interval, showing that the proportion of closed 4-paths with additional edges is much higher than expected. Intuitively, as the southern women network is a social network, one would assume that any of the 18 women would rather attend an event with friends than by herself. Our results confirm the assumption that three women tend to cluster if they are already connected to each other by at least one event. 

\begin{table}[h]
	\caption{The local clustering coefficients of the 18 women and their driving scores. The 7 women who were identified to drive the clustering behaviour of the whole network are printed in bold.}\label{localWomen}
	\centering
	\scriptsize
	\begin{tabular}{llllll}
		\toprule
		woman $i$ & $cc_{(i,0)}$ & $cc_{(i,1)}$ & $cc_{(i,2)}$ & $cc_{(i,3)}$ & $ds_{(i)}$\\
		\midrule
		{\bf Evelyn} &  0.3957 \hfill $\downarrow$ & 0.6986 \hfill $\uparrow$&  0.6732  \hfill $\uparrow$& 0.6545  \hfill $\uparrow$ & 0.4083\\
		{\bf Laura} &  0.4468 \hfill $\downarrow$ &  0.6610 \hfill $\uparrow$&  0.7218  \hfill $\uparrow$&  0.7364  \hfill $\uparrow$ & 0.4179\\
		{\bf Theresa} &  0.0619 \hfill $\downarrow$ & 0.7228 \hfill $\uparrow$& 0.7951  \hfill $\uparrow$& 0.6667  \hfill $\uparrow$ & 0.6092\\
		{\bf Brenda} & 0.3455 \hfill $\downarrow$ & 0.656   \hfill $\uparrow$& 0.7241  \hfill $\uparrow$& 0.7565  \hfill $\uparrow$ & 0.4633\\
		Charlotte & 1  \hfill $\uparrow$        & 0.84     \hfill $\uparrow$& 0.6093  \hfill $\uparrow$& 0.6  \hfill $\uparrow$& 0.0962\\
		Frances & 0.6667 \hfill $\uparrow$ & 0.684  \hfill $\uparrow$ & 0.5164  \hfill $\uparrow$& 0.7742  \hfill $\uparrow$& 0.2626\\
		{\bf Eleanor} & 0.5094 \hfill $\downarrow$ & 0.662  \hfill $\uparrow$ & 0.6302  \hfill $\uparrow$& 0.6234  \hfill $\uparrow$ & 0.3133\\
		Pearl & 0.4074 \hfill $\downarrow$ & 0.6931\hfill $\uparrow$ & 0.4278  \hfill $=$& 0.0652  \hfill $\downarrow$& -0.0254\\
		{\bf Ruth} & 0.2869 \hfill $\downarrow$ & 0.697  \hfill $\uparrow$ & 0.6254  \hfill $\uparrow$& 0.3704  \hfill $=$ & 0.3248\\
		Verne & 0.3778 \hfill $\downarrow$ & 0.613\hfill $\uparrow$   & 0.6188  \hfill $\uparrow$& 0.3429  \hfill $=$ & 0.2253\\
		Myrna & 0.6735 \hfill $\uparrow$& 0.5221 \hfill $\downarrow$& 0.504    \hfill $\uparrow$& 0.4615  \hfill $\uparrow$ & 0.04978\\
		Katherine & 0.7260 \hfill $\uparrow$ & 0.569   \hfill $\uparrow$& 0.5572  \hfill $\uparrow$& 0.5254 \hfill $\uparrow$ & 0.0822\\
		{\bf Sylvia} & 0.3395 \hfill $\downarrow$ & 0.6694 \hfill $\uparrow$& 0.653    \hfill $\uparrow$& 0.5444 \hfill $\uparrow$ & 0.3646\\
		Nora & 0.7185 \hfill $\uparrow$ & 0.7555 \hfill $\uparrow$& 0.4021  \hfill $\downarrow$& 0.5238  \hfill $\uparrow$ & 0.1247\\
		Helen & 0.7143 \hfill $\uparrow$ & 0.6273 \hfill $\uparrow$& 0.4703  \hfill $\uparrow$& 0.375 \hfill $=$ & 0.0308\\
		Dorothy & 0.4667 \hfill $\downarrow$ & 0.4557 \hfill $\downarrow$& 0.163    \hfill $\downarrow$& 0  \hfill $\downarrow$ & -0.3793\\
		Olivia & 1 \hfill $\uparrow$ & 0.3103 \hfill $\downarrow$& 0  \hfill $\downarrow$& 0 \hfill $\downarrow$ & -0.8607\\
		Flora & 1 \hfill $\uparrow$ & 0.3103 \hfill $\downarrow$& 0  \hfill $\downarrow$& 0  \hfill $\downarrow$&  -0.8607\\
		\bottomrule    
	\end{tabular}
\end{table}

The global driving score of the southern women networks is $ds_{(global)} = 0.297$. The local clustering coefficients of the southern women network, together with the respective driving scores are displayed in Table \ref{localWomen}. The arrows next to the entries in the table, indicate if the local clustering coefficient is higher or lower than the respective global clustering coefficient. 

The driving scores of the women, reveal that Evelyn, Laura, Theresa, Brenda, Eleanor, Ruth and Sylvia drive the clustering behaviour of the whole network. A woman drives the clustering behaviour if her driving score lies above the global driving score of the network. All nodes with a negative score drive against the overall clustering behaviour.  

We repeat the analysis for the secondary node set that represents the 14 events. Table \ref{resultsWomenPassive} shows the clustering coefficients of the southern women network with respect to the events. 

\begin{table}[h]
		\caption{The four clustering coefficients of the southern women network with respect to the passive node set of events and the average clustering coefficients of 100 randomly generated networks with their 95\% confidence interval.} \label{resultsWomenPassive}
	\scriptsize
	\centering
	\begin{tabular}{lll}
		\toprule
		& Southern Women network & random networks\\
		\midrule
		$cc_{(0)}$ & 0.3578 & [0.7164, 0.7412]\\
		$cc_{(1)}$ & 0.597 & [0.6272, 0.65]\\
		$cc_{(2)}$ & 0.8556 & [0.4871, 0.5209]\\
		$cc_{(3)}$ & 0.7903 & [0.422, 0.4757] \\
		\bottomrule
	\end{tabular}
\end{table} 

Again, none of the four clustering coefficients lie within the 95\% confidence interval of the randomly generated networks. The coefficients $cc_{(0)}$ and $cc_{(1)}$ lie below the lower bound of the respective confidence interval whereas the $cc_{(2)}$ and $cc_{(3)}$ lie above the interval. 

Calculation of the driving scores of the 14 events, displayed in Table \ref{localWomenPassive}, shows that events 3, 5, 6 and 8 drive the clustering behaviour of the network.

\begin{table}[h]
	\caption{The local clustering coefficients of the 14 events and their driving scores. The events that were identified to drive the clustering behaviour of the whole network are printed in bold. The global driving score with respect to the events equals 0.4756.} \label{localWomenPassive}
	\centering
	\scriptsize
	\begin{tabular}{llllll}
		\toprule
		event $i$ & $cc_{(i,0)}$ & $cc_{(i,1)}$ & $cc_{(i,2)}$ & $cc_{(i,3)}$ & $ds_{(i)}$\\
		\midrule
		1 &  1 \hfill $\uparrow$&  0.9556 \hfill $\uparrow$ &  0.7714 \hfill $\downarrow$& 0.6 \hfill $\downarrow$& -0.2659\\
		2 & 0.8 \hfill $\uparrow$&  0.9574 \hfill $\uparrow$&  0.8571 \hfill $\uparrow$&  0.5143 \hfill $\downarrow$ & -0.0785\\
		{\bf 3} &  0.3043 \hfill $\downarrow$&  0.7113 \hfill $\uparrow$& 0.9727  \hfill $\uparrow$& 0.8824  \hfill $\uparrow$ & 0.5281\\
		4 & 0.9 \hfill $\uparrow$ & 0.9529   \hfill $\uparrow$& 0.8803  \hfill $\uparrow$&  0.6427  \hfill $\uparrow$ & -0.0976\\
		{\bf 5} & 0.2545  \hfill $\downarrow$        & 0.7952     \hfill $\uparrow$& 0.9895  \hfill $\uparrow$& 0.9029  \hfill $\uparrow$ & 0.505\\
		{\bf 6} & 0.3421 \hfill $\downarrow$ &  0.5482  \hfill $\downarrow$ & 0.8913  \hfill $\uparrow$&  0.8791  \hfill $\uparrow$ & 0.5584\\
		7 & 0.3195 \hfill $\downarrow$ & 0.6965 \hfill $\uparrow$ & 0.8165  \hfill $\uparrow$& 0.7051  \hfill $\uparrow$ & 0.374\\
		{\bf 8} & 0.38 \hfill $\downarrow$ & 0.5918 \hfill $\downarrow$ & 0.9429  \hfill $\uparrow$&  0.8672  \hfill $\uparrow$ & 0.5489\\
		9 & 0.3062 \hfill $\downarrow$ & 0.6823  \hfill $\uparrow$ & 0.7968  \hfill $\uparrow$& 0.6923  \hfill $\uparrow$ & 0.3727\\
		10 & 0.48 \hfill $\downarrow$ & 0.7023\hfill $\uparrow$   & 0.7891  \hfill $\uparrow$& 0.8049  \hfill $\uparrow$ & 0.3465\\
		11 & 1 \hfill $\uparrow$& 0.7949 \hfill $\uparrow$& 0.1    \hfill $\downarrow$& 0  \hfill $\downarrow$ & -0.8085\\
		12 & 0.3889 \hfill $\downarrow$ & 0.7348   \hfill $\uparrow$& 0.8187  \hfill $\uparrow$& 0.875 \hfill $\uparrow$ & 0.4019\\
		13 & 1 \hfill $\uparrow$ & 0.6098 \hfill $\downarrow$& 0.5323    \hfill $\uparrow$& 0.6923 \hfill $\uparrow$ & -0.114\\
		14 &1 \hfill $\uparrow$ & 0.6098 \hfill $\downarrow$& 0.5323    \hfill $\uparrow$& 0.6923 \hfill $\uparrow$ & -0.114 \\
		\bottomrule
	\end{tabular}
\end{table}

\subsection{Discussion}\label{subsec:discussion}

\begin{figure}[h]
	\centering
	\includegraphics[width=0.8\textwidth]{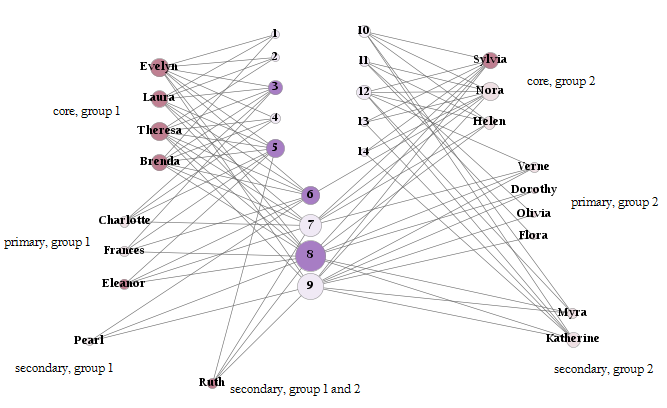}%
	\caption{This figure shows the southern women network. The two groups and their core, primary  and secondary members that were identified by Davis, Gardner and Gardner \cite{dav41} are labelled. The darker shaded nodes are driving the clustering behaviour of the network, as identified by our analysis. The size of the nodes corresponds to their degrees.}\label{core}
\end{figure}

The first analysis of the southern women dataset was carried out by Davis, Gardner and Gardner \cite{dav41} in the form of interviews, with the aim to categorise the 18 women into groups. They found two different groups that were further divided into core, primary and secondary members. Figure \ref{core} shows the southern women network with the two groups identified in \cite{dav41}. Our analysis found that all the core women of the first group are influential as well as one core woman of the second group. Interestingly, our results show that Eleanor and Ruth should also be considered as important. Both attended only four events, however, these events were also attended by members from both groups. This observation indicates that Eleanor and Ruth are an important connection between the two groups. Davis, Gardner and Gardner also found that Ruth had some affiliation with both groups. Clearly our clustering coefficient identifies important nodes across the communities that were identified by Davis, Gardner and Gardner. Our analysis shows that the importance of a woman does not depend on her degree. For instance, if Ruth, who has a low degree, is removed from the network, information would spread less easily between the two groups.

Dorothy, Olivia, Flora and Pearl received negative driving scores. This result is consistent with the results presented in \cite{bon78}, \cite{dor79} and \cite{eve93} which found that Dorothy, Olivia, Flora and Pearl were not associated with any of the groups.  

The events that our analysis identified seem to be important not only because they have a high degree, but also because they were attended by women from both groups and hence act as a connection between the two communities. 

\section{The Noordin Top Terrorist Network}\label{sec:Noordin}
We now investigate a subset of the Noordin Top Terrorist network \cite{eve12}. This particular subset models the attendance of 26 members of the terrorist network at 20 different meetings. The subset contains a total of 64 connections between members and meetings. Table \ref{resultsNoordin} shows the clustering coefficients of the members of the terrorist ring.

\begin{table}[h]
		\caption{The four clustering coefficients of the terrorist network and the average clustering coefficients of 100 randomly generated networks with their 95\% confidence interval with respect to the members of the terrorist ring. }\label{resultsNoordin}
	\scriptsize
	\centering
	\begin{tabular}{lll}
		\toprule
		& Noordin Top Terrorist Network & random network\\
		\midrule
		$cc_{(0)}$ & 0.0303 & [0.1768, 0.1973]\\
		$cc_{(1)}$ & 0.1108 & [0.0542, 0.0676]\\
		$cc_{(2)}$ & 0.2 & [0.0199, 0.0376]\\
		$cc_{(3)}$ & 0 & [0, 0.0148] \\
		\bottomrule
	\end{tabular}
\end{table}

The clustering coefficient $cc_{(0)}$ of the members lies below the confidence interval, whereas $cc_{(1)}$ and $cc _{(2)}$ lie above the interval. The clustering coefficient $cc_{(3)}$, however, lies within the 95\% confidence interval.

In the terrorist network, the proportion of 4-paths that are closed and form a highly connected 6-cycle is much higher than in a random network ($cc_{(2)} = 0.2$). As in the southern women network, it seems that three members of the terrorist ring would cluster if they were already connected through at least one previous meeting. The results from the southern women network can be explained by the underlying friendship network. In case of the terrorist network, it is rather unlikely that the members decided which meetings to attend, based on friendships to other members. Examination of the local clustering coefficients reveal the important members (see Table \ref{localNoordin}).

\begin{table}[h]
	\caption{This table shows the local clustering coefficients of the 26 members of the Noordin terrorist network. The four members that were identified to drive the clustering behaviour of the whole network are printed in bold. The global driving score with respect to the members equals 0.2692. }\label{localNoordin}
	\centering
	\scriptsize
	\begin{tabular}{p{2cm}lllll}
		\toprule
		member $i$ & $cc_{(i,0)}$ & $cc_{(i,1)}$ & $cc_{(i,2)}$ & $cc_{(i,3)}$& $ds_{(i)}$\\
		\midrule
		Abdullah Sunata & n/a & n/a & n/a & n/a & n/a\\
		Abu Dujanah & n/a & 0 \hfill $\downarrow$& 0.1667 \hfill $\uparrow$& 0\hfill $=$ & 0.0473\\
		Abu Fida & 0.1667 \hfill $\downarrow$& 0.1333 \hfill $\uparrow$& 0 \hfill $\downarrow$&  n/a & -0.2713\\
		Adung &n/a&n/a&n/a&n/a & n/a\\
		{\bf Ahmad Rofiq Ridho} & 0.0408 \hfill $\downarrow$ & 0.1818 \hfill $\uparrow$& 0.2414 \hfill $\uparrow$& 0\hfill $=$ & 0.5324\\
		Akram &n/a&n/a&n/a&n/a &n/a\\
		Asep Jaja &n/a&n/a&n/a&n/a &n/a\\
		{\bf Azhari Husin} &  0 \hfill $\downarrow$& 0.0842 \hfill $\uparrow$&  0.2857 \hfill $\uparrow$& 0 \hfill $=$ & 0.5723\\
		Cholily &n/a&n/a&n/a&n/a &n/a\\
		Heri Sigu Samboja &n/a&n/a&n/a&n/a &n/a\\
		Imam Bukhori &n/a&n/a&n/a&n/a &n/a\\
		Ismail &n/a&n/a&n/a&n/a &n/a\\
		Iwan Dharmawan & 0.1429 \hfill $\downarrow$& 0.2609 \hfill $\uparrow$& 0 \hfill $\downarrow$& n/a & -0.1836\\
		Jabir &n/a&n/a&n/a&n/a &n/a\\
		Joko Triharmanto &n/a&n/a&n/a&n/a &n/a\\
		Misno &n/a&n/a&n/a&n/a &n/a\\
		Mohamed Saifuddin & 0 \hfill $\downarrow$& 0 \hfill $\downarrow$& n/a & n/a & 0 \\
		{\bf Noordin Mohammed Top} & 0.0141 \hfill $\downarrow$& 0.124 \hfill $\uparrow$& 0.2079 \hfill $\uparrow$& 0\hfill $=$ & 0.5442\\
		Purnama Putra & 0.1429 \hfill $\downarrow$& 0.3333 \hfill $\uparrow$& 0.3333 \hfill $\uparrow$& 0\hfill $=$ & 0.46\\
		Qotadah & n/a & 0 \hfill $\downarrow$& 0.1667 \hfill $\uparrow$& 0\hfill $=$ & 0.0473\\
		Saptono &n/a&n/a&n/a&n/a &n/a\\
		Son Hadi & 0.1667 \hfill $\downarrow$& 0 \hfill $\downarrow$& n/a& n/a & -0.4454\\
		Suramto &n/a&n/a&n/a&n/a &n/a\\
		Ubeid &n/a&n/a&n/a&n/a &n/a\\
		Urwah & 0.2 \hfill $=$& 0.3333 \hfill $\uparrow$& 0 \hfill $\downarrow$& n/a & -0.2419\\
		Usman bin Sef & 0 \hfill $\downarrow$& 0 \hfill $\downarrow$& n/a & n/a & 0\\
		\bottomrule
	\end{tabular}
\end{table}

The driving scores clearly show the influential nodes who are driving the clustering behaviour of the network are Ahmad Rofiq Ridho, Azhari Husin and Noordin Mohammed Top. Noordin Mohammed Top and Azhari Husin worked together to plan the terrorist attacks, with Noordin Mohammed Top financing the attacks and Azhari Husin being in charge of building the bombs \cite{bbc14}. Ahmad Rofiq Ridho was acting as a communicator between the members \cite{eve12}. Purnama Putra also received a very high driving score and was taking the role of a communicator similar to that Ahmad Rofiq Ridho. 

The driving scores of the secondary node set revealed that meetings 16 and 18 are driving the clustering behaviour. Unfortunately, we do not have enough information about the meetings or the terrorist ring to explain our results. 

\begin{figure}[h]
	\centering
	\includegraphics[width=0.8\textwidth]{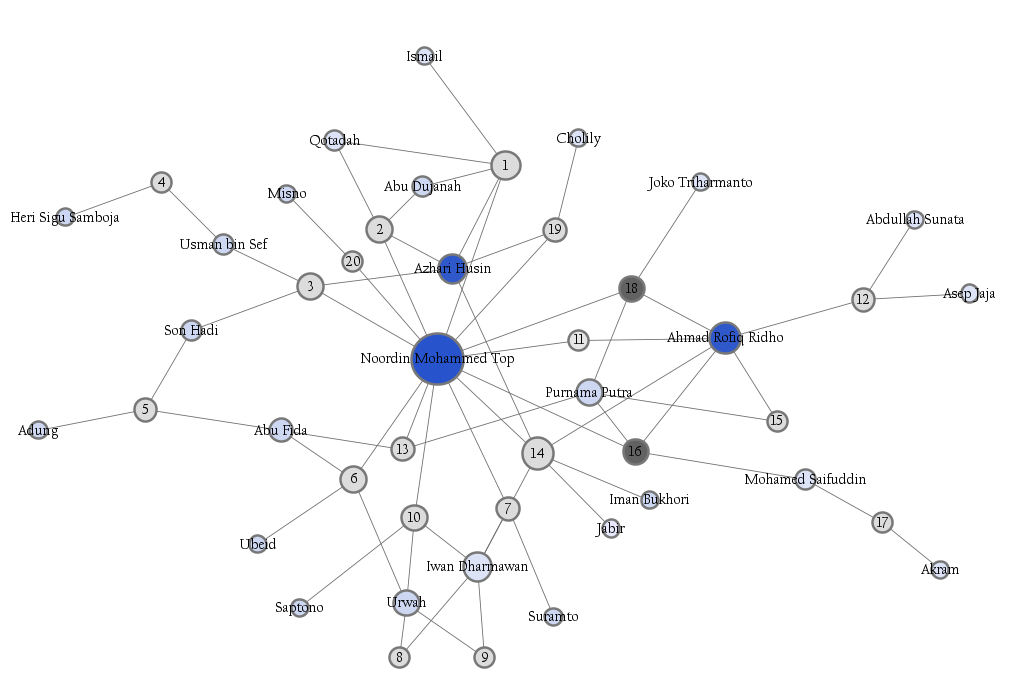}%
	\caption{This figure shows the Noordin Top terrorist network. The darker shaded nodes are driving the clustering behaviour of the network, as identified by our analysis. The size of the nodes corresponds to their degrees.}\label{im:noordin}
\end{figure} 

Figure \ref{im:noordin} shows the terrorist network. The influential nodes have a darker shading. Even though Noordin Top is one of the most influential members in the network and also has the highest degree, in general the importance of a member does not depend on its degree. 

\section{Conclusion and Future Work}\label{sec:conclusion}
Many real world networks are bipartite. However, not every network measure can be directly applied to this type of network \cite{lat08}. In order to analyse bipartite networks, one can either project the network to a one-mode network or redefine those networks measure that are not suitable for the analysis of bipartite networks.

A measure that has received much recent interest is the clustering coefficient. However, the existing methods do not consider the different structures that a bipartite cluster may have.  

This paper showed that it is important to distinguish between different types of 6-cycles that are identified by the number of additional edges that connect nodes within the cycle. Ignoring additional edges results in an over-count of 6-cycles. We showed that the formation of the different types of 6-cycles depends on the network. For instance, in a network where primary nodes may connect to secondary nodes at any point in time, a sparsely connected 6-cycle can originate from an unconnected 6-cycle. This is not possible in a time dependent network where any 6-cycle can only originate from a 4-path. We defined four clustering coefficients that correspond to the different types of 6-cycles. 

Applying the four clustering coefficients to real world networks gives valuable insight into how clusters are structured and how they form. The driving scores of the individual nodes in a network revealed those nodes that are driving the clustering behaviour and have influence on the network structure. Previous analyses supports the results we obtained in Section \ref{sec:women}. 

Here, we tested our approach on relatively small networks. The next step would be to test the performance of our method on large-scale bipartite networks. 

Considering time stamps of a network, if they are available, could give further insight into the network of interest and make a more dynamic analysis of the network possible. The clustering coefficients that were introduced in this paper can only be applied to time dependent bipartite networks. Further work needs to be done on other types of networks in which 6-cycles are formed differently. We will also focus future work on finding whole communities in bipartite networks by applying the clustering coefficients proposed in this paper. 

\bibliography{bibliography}

\end{document}